  \documentstyle[aas2pp4,epsf]{article}

\newcommand{\Ell}{E_\parallel}      
\newcommand{\rhoGJ}{\rho_{{\rm GJ}}}  
\newcommand{\sgT}{\sigma_{\rm T}}  
\newcommand{\sgP}{\sigma_{\rm p}}  
\newcommand{\rlc}{\varpi_{\rm LC}} 
\newcommand{\Ex}{\epsilon_{\rm x}} 
\newcommand{\Eg}{\epsilon_\gamma}  

\lefthead{Hirotani and Shibata}
\righthead{}

\begin{document}

\title{Electrodynamic Structure of an Outer--Gap Accelerator:
       Gamma-Ray Emission from the Crab Pulsar}
\author{K. Hirotani}
\affil{National Astronomical Observatory, 
       Mitaka, Tokyo 181-8588, Japan\\
       hirotani@hotaka.mtk.nao.ac.jp}
\and
\author{S. Shibata}
\affil{Department of Physics, Yamagata University,
       Yamagata 990-8560, Japan\\
       shibata@sci.kj.yamagata-u.ac.jp}

\begin{abstract}
We investigate a stationary pair production cascade 
in the outer magnetosphere of a spinning neutron star.
The charge depletion due to a global current,
causes a large electric field along the magnetic field lines.
Migratory electrons and/or positrons are accelerated by this field
to radiate curvature gamma-rays, 
some of which collide with the X-rays to materialize as pairs in the gap.
The replenished charges partially screen the electric field, 
which is self-consistently solved together with the distribution functions
of particles and gamma-rays.
If no current is injected at neither of the boundaries of the accelerator,
the gap is located around the so-called null surface,
where the local Goldreich-Julian charge density vanishes.
However, we first find that the gap position shifts outwards (or inwards)
when particles are injected at the inner (or outer) boundary.
Applying the theory to the Crab pulsar, 
we demonstrate that 
the pulsed TeV flux does not exceed the observational upper limit 
for moderate infrared photon density 
and that the gap should be located near to or outside of the null surface
so that the observed spectrum of pulsed GeV fluxes may be 
emitted via curvature process.
\end{abstract}



\section{Introduction}

The EGRET experiment on the Compton Gamma Ray Observatory
has detected pulsed signals from seven rotation-powered pulsars
(e.g., for Crab, Nolan et al. 1993, Fierro et al. 1998).
The modulation of the $\gamma$-ray light curves at GeV energies 
testifies to the production of $\gamma$-ray radiation in the pulsar 
magnetospheres either at the polar cap 
(Harding, Tademaru, \& Esposito 1978; Daugherty \& Harding 1982, 1996;
 Sturner, Dermer, \& Michel 1995;
 Shibata, Miyazaki, \& Takahara 1998),
or at the vacuum gaps in the outer magnetosphere
(Cheng, Ho, \& Ruderman 1986a,b, hereafter CHR;
 Chiang \& Romani 1992, 1994; Romani and Yadigaroglu 1995;
 Romani 1996; Zhang \& Cheng 1997, heafter ZC97).
Effective $\gamma$-ray production in a pulsar magnetosphere
may be extended to the very high energy (VHE) region above 
100 GeV as well;
however, the predictions of fluxes by the current models of 
$\gamma$-ray pulsars are not sufficiently conclusive
(e.g., Cheng 1994).
Whether or not the spectra of $\gamma$-ray pulsars continue up to the
VHE region is a question which remains one of the 
interesting issues of high-energy astrophysics.

In the VHE region,
positive detections of radiation at a high confidence 
level have been reported from the direction of the Crab pulsar
(Nel et al. 1993).
However, as for {\it pulsed} TeV radiation,
only the upper limits have been, as a rule, obtained.
If the VHE emission originates the pulsar magnetosphere,
a significant fraction of them can be expected to show pulsation.
Therefore, the lack of {\it pulsed} TeV emissions provides a
severe constraint on the modeling of particle acceleration zones
in a pulsar magnetosphere.

In fact, in CHR picture,
the magnetosphere should be optically thick for pair--production
in order to reduce the TeV flux to an unobserved level 
by absorption.
This in turn requires very high luminosities of infrared photons.
However, the required IR fluxes are generally orders of magnitude
larger than the observed values (Usov 1994).
We are therefore motivated by the need to contrive an outer--gap model
which produces less TeV emission with a moderate infrared luminosity.

High-energy emission from a pulsar magnetosphere,
in fact, crucially depends on the acceleration electric field, 
$\Ell$, along the magnetic field lines.
It was Hirotani and Shibata (1999a,b,c; hereafter Papers I, II, III),
and Hirotani (2000a,b,c; hereafter Papers IV, V, VI)
who first solved the spatial distribution of $\Ell$ 
together with particle and $\gamma$-ray distribution functions.
They demonstrated that 
a stationary gap is formed around the null surface at which the 
local Goldreich--Julian charge density, 
\begin{equation}
  \rho_{\rm GJ}= \frac{\Omega B_z}{2\pi c},
  \label{eq:def_rhoGJ}
\end{equation}
vanishes,
where $B_z$ is the component of the magnetic field along 
the rotation axis,
$\Omega$ the angular frequency of the neutron star,
and $c$ the speed of light.
Equation (\ref{eq:def_rhoGJ}) is valid unless the gap is 
located close to the light cylinder,
of which distance from the rotation axis is
given by $ \rlc= c / \Omega$.
In this letter, 
we develop the method presented in Paper VI,
by investigating the case when particles flow
into the gap from the inner or the outer boundaries.

In the next two sections, we describe the physical processes of pair 
production cascade and the resultant $\gamma$-ray emission.
We then apply the theory to the Crab pulsar 
and present the expected $\gamma$-ray spectra in \S~4.
In the final section, we compare the results with ZC97.

\section{Basic Equations and Boundary Conditions}
\label{sect2}

Let us first consider the Poission equation for the electrostatic 
potential, $\Phi$.
Neglecting relativistic effects,
and assuming that typical transfield thickness of the gap, $D_\perp$, 
is greater than or comparable with the longitudinal gap width, $W$,
we can reduce the Poisson equation into the form (Paper~VI)
\begin{equation}
  -\frac{d^2}{ds^2} \Phi 
    = 4 \pi e (N_+ -N_- -\rhoGJ/e),
  \label{Poission_1}
\end{equation}
where $N_+$ and $N_-$ designate the positronic and electronic densities,
respectively,
$e$ the magnitude of the charge on an electron,
and $s$ the length along the last-open fieldline.

It is convenient to non-dimensionalize the length scales 
by a typical Debey scale length $c/\omega_{\rm p}$, where
\begin{equation}
  \omega_{\rm p} = \sqrt{ \frac{4\pi e^2}{m_{\rm e}}
	                  \frac{\Omega B_{\rm c}}{2\pi ce} };
  \label{eq:def-omegap}
\end{equation}
$B_{\rm c}$ represents the magnetic field strength at the gap center.
The dimensionless coordinate variable then becomes
\begin{equation}
  \xi \equiv (\omega_{\rm p}/c) s.
  \label{eq:def-xi}
\end{equation}
By using such dimensionless quantities, we can rewrite
the Poisson equation into
\begin{equation}
  E_\parallel = -\frac{d\varphi}{d\xi},
  \label{eq:basic-1}
\end{equation}
\begin{equation}
  \frac{dE_\parallel}{d\xi}
  = \frac{B(\xi)}{B_{\rm c}} \left[ n_+(\xi) - n_-(\xi) \right]
    + \frac{B_z(\xi)}{B_{\rm c}}
  \label{eq:basic-2}
\end{equation}
where $ \varphi(\xi) \equiv e\Phi(s)/(m_{\rm e}c^2)$;
the particle densities per unit flux tube are defined by
\begin{equation}
  n_\pm(\xi) \equiv 
    \frac{2\pi ce}{\Omega} \frac{N_\pm(s)}{B(s)}.
  \label{eq:def-n}
\end{equation}
We evaluate $B_z/B$ at each point along the last-open field line,
by using the Newtonian dipole field.

We next consider the continuity equations for the particles.
Assuming that both electrostatic and 
curvature-radiation-reaction forces cancel out each other,
we obtain the following continuity equations
\begin{equation}
  \pm B \frac{d}{ds}\left( \frac{N_\pm}{B} \right)
  = \frac{1}{c} \int_{0}^\infty d\epsilon_\gamma \, 
    [ \eta_{\rm p+} G_+   +\eta_{\rm p-} G_- ],
  \label{eq:cont-eq}
\end{equation}
where $G_\pm(s,\epsilon_\gamma)$ are the distribution functions of
$\gamma$-ray photons having momentum $\pm m_{\rm e}c \epsilon_\gamma$ 
along the poloidal field line.
Since the electric field is assumed to be positive in the gap,
$e^+$'s (or $e^-$'s) migrate outwards (or inwards).
The pair production redistribution
functions, $\eta_\pm$, are defined as
\begin{equation}
  \eta_{{\rm p}\pm}(\Eg)
  = (1-\mu_{\rm c}) \frac{c}{\omega_{\rm p}}
     \int_{\epsilon_{\rm th}}^\infty d\epsilon_{\rm x}
     \frac{dN_{\rm x}}{d\Ex} 
     \sgP(\Eg,\Ex,\mu_{\rm c}),
  \label{eq:def_etap_0}
\end{equation}
where $\sgP$ is the pair-production cross section and
$\cos^{-1}\mu_{\rm c}$ refers to the collision angle between 
the $\gamma$-rays and the X-rays 
(see Paper~VI for more details about eq.~[\ref{eq:def_etap_0}]).

Let us introduce the following dimensionless $\gamma$-ray 
densities in the dimensionless energy interval
between $\beta_{i-1}$ and $\beta_i$:
\begin{equation}
  g_\pm{}^i(\xi) \equiv 
    \frac{2\pi ce}{\Omega B_{\rm c}}
    \int_{\beta_{i-1}}^{\beta_i} d\epsilon_\gamma G_\pm(s,\epsilon_\gamma).
  \label{eq:def-g}
\end{equation}
In this letter, we set $\beta_0=10^2$,
which corresponds to the lowest $\gamma$-ray energy, $51.1$ MeV.
We divide the $\gamma$-ray spectra into $9$ energy bins 
and put
$\beta_1= 10^{2.5}$, 
$\beta_2= 10^3$, 
$\beta_3= 10^{3.5}$, 
$\beta_4= 10^4$, 
$\beta_5= 10^{4.5}$, 
$\beta_6= 10^{4.75}$, 
$\beta_7= 10^5$. 
$\beta_8= 10^{5.25}$, and
$\beta_9= 10^{5.5}$.

We can now rewrite equation
(\ref{eq:cont-eq}) into 
\begin{equation}
  \frac{dn_\pm}{d\xi} = 
    \pm \frac{B_{\rm c}}{B(\xi)}
        \sum_{i=1}^{9} [ \eta_{\rm p+}{}^i g_+^i(\xi)
                        +\eta_{\rm p-}{}^i g_-^i(\xi)],
  \label{eq:basic-3}
\end{equation}
where $\eta_{{\rm p}\pm}^i$ 
are evaluated at the central energy in each bin. 

A combination of equations (\ref{eq:basic-3}) 
gives the current conservation law,
\begin{equation}
  j_{\rm tot} \equiv n_+(\xi) + n_-(\xi) = {\rm constant \ for \ } \xi.
  \label{eq:consv}
\end{equation}
When $j_{\rm tot}=1.0$, 
the current density per unit flux tube
equals the Goldreich--Julian value, $\Omega / (2\pi)$.

Unlike the charged particles,
$\gamma$-rays do not propagate along the local magnetic field lines.
However, to avoid complications, we simply assume 
that the outwardly (or inwardly) propagating $\gamma$-rays
dilate (or constrict) at the same rate with the magnetic field.
This assumption gives a good estimate 
when $W \ll \rlc$ holds.
We then obtain (Paper~VI)
\begin{eqnarray}   
  \frac{d}{d\xi} g_\pm{}^i(\xi)
     = \frac{d}{d\xi}\left( \ln B \right)
       \mp \eta_{{\rm p}\pm}{}^i g_\pm{}^i
       \pm \eta_{\rm c}^i \frac{B(\xi)}{B_{\rm c}} n_\pm(\xi),
  \label{eq:basic-5}
\end{eqnarray}   
where $i=1,2,\cdot\cdot\cdot,m$ ($m=9$) and 
\begin{eqnarray}
  \eta_{\rm c}^i 
  &\equiv& \frac{\sqrt{3}e^2\Gamma}{\omega_{\rm p}hR_{\rm c}}
           \int_{\beta_{i-1} / \epsilon_{\rm c}}
               ^{\beta_i     / \epsilon_{\rm c}}
            ds \int_s^\infty K_{\frac53}(t)dt, 
  \label{eq:etaCi}
\end{eqnarray}
where $K_{5/3}$ refers to the modified Bessel function of $5/3$ order.

Equating the electric force $e \vert d\Phi / dx \vert$ and the
radiation reaction force,
we obtain the saturated Lorentz factor at each point as 
\begin{equation}
  \Gamma_{\rm sat} 
   = \left( \frac{3 R_{\rm c}{}^2}{2e} 
		    \left| \frac{d\Phi}{dx} \right|
                  + 1 
     \right)^{1/4};
  \label{eq:saturated}
\end{equation}
we compute the curvature radius $R_{\rm c}$ 
at a point for a Newtonian dipole magnetic field.
Since the maximum of $\vert d\Phi/dx \vert$ and the potential drop
are roughly proportional to $W^2$ and $W^3$, respectively (Paper~V),
the particles become unsaturated for very small $W$.
To avoid an overestimation in such cases, 
we compute $\Gamma$ by 
\begin{equation}
  \frac{1}{\Gamma}
  = \sqrt{ \frac{1}{\Gamma_{\rm sat}{}^2}
          +\frac{1}{\varphi^2(\xi_2)}
         },
  \label{eq:terminal}
\end{equation}
where $\varphi(\xi_2)$ represents the maximum attainable Lorentz factor.

\subsection{Boundary Conditions}
\label{sec:BD}

To solve the differential equations
(\ref{eq:basic-1}), (\ref{eq:basic-2}), (\ref{eq:basic-3}), 
and (\ref{eq:basic-5}),
we must impose boundary conditions.
At the {\it inner} (starward) boundary
($\xi= \xi_1$), we impose (Paper~VI)
\begin{equation}
  E_\parallel(\xi_1)=0,
  \quad
  \varphi(\xi_1) = 0,
  \label{eq:BD-1}
\end{equation}
\begin{equation}
  g_+{}^i(\xi_1)=0  \quad (i=1,2,\cdot\cdot\cdot,9).
  \label{eq:BD-3}
\end{equation}
Since positrons may flow into the gap at $\xi=\xi_1$
as a part of the global current pattern in the magnetosphere,
we denote the positronic current per unit flux tube at $\xi=\xi_1$ as
\begin{equation}
  n_+(\xi_1)= j_1,
  \label{eq:BD-4}
\end{equation}
which yields (eq.~[\ref{eq:consv}])
\begin{equation}
  n_-(\xi_1)= j_{\rm tot}-j_1.
  \label{eq:BD-5}
\end{equation}

At the {\it outer} boundary ($\xi=\xi_2$), we impose
\begin{equation}
  E_\parallel(\xi_2)=0, \quad
  g_-{}^i(\xi_2)=0 \quad (i=1,2,\cdot\cdot\cdot,9),
  \label{eq:BD-7}
\end{equation}
\begin{equation}
  n_-(\xi_2)= j_2.
  \label{eq:BD-8}
\end{equation}

The current density created in the gap per unit flux tube
can be expressed as
\begin{equation}
  j_{\rm gap}= j_{\rm tot} -j_1 -j_2.
  \label{eq:Jgap}
\end{equation}
We adopt $j_{\rm gap}$, $j_1$, and $j_2$
as the free parameters.

We have totally $24$ boundary conditions 
(\ref{eq:BD-1})--(\ref{eq:BD-8})
for $22$ unknown functions
$\Phi$, $E_\parallel$,
$n_\pm$, 
$g_\pm{}^i$ ($i=1, 2, \cdot\cdot\cdot, 9$).
Thus two extra boundary conditions must be compensated 
by making the positions of the boundaries $\xi_1$ and $\xi_2$ be free.
The two free boundaries appear because $E_\parallel=0$ is imposed at 
{\it both} the boundaries and because $j_{\rm gap}$ is externally imposed.
In other words, the gap boundaries ($\xi_1$ and $\xi_2$) shift,
if $j_1$ and/or $j_2$ varies.

\section{TeV Spectra}
\label{sec:TeV_spc}

For simplicity, assume that the IR field are homogeneous
and isotropic within the radius $\rlc$.
Interpolating the pulsed fluxes in radio, near IR, and optical bands
from the Crab pulsar
(Moffett and Hankins 1996; Percival et al. 1993; Eikenberry et al. 1997),
we obtain (Paper~V) 
\begin{equation}
  \frac{dN_{\rm IR}}{d\epsilon_{\rm IR}}
    = 1.5 \times 10^{17} d^2 
      \left(\frac{r_0}{\rlc}\right)^{-2}
      \epsilon_{\rm IR}{}^{-0.88},
  \label{eq:IR_Crab}
\end{equation}
where $\epsilon_{\rm IR} m_{\rm e}c^2$ 
refers to the IR photon energy,
and 
$\epsilon_{\rm IR,min} < \epsilon < \epsilon_{\rm IR,max}$.
We adopt 
$\epsilon_{\rm IR,min}= 10^{-6}$ and
$\epsilon_{\rm IR,min}= 10^{-2}$;
the results do not depend on these cut-off energies very much.

If an electron or a positron is migrating 
with Lorentz factor $\Gamma \gg 1$ in an isotropic photon field,
it upscatters the soft photons to produce
the following number spectrum of $\gamma$-rays
(Blumenthal \& Gould 1970):
\begin{eqnarray}
  \frac{d^2 N}{dtd\Eg}
  &=& \frac34 \sgT \frac{c}{\Gamma^2}
      \frac{dN_{\rm IR}}{d\epsilon_{\rm IR}}
      \frac{d\epsilon_{\rm IR}}{\epsilon_{\rm IR}}
  \nonumber \\
  & & \hspace{-2.0 truecm}
      \times
      \left[ 2q \ln q +(1+2q)(1-q)
            +\frac{(Qq)^2(1-q)}{2(1+Qq)}
      \right],
  \label{eq:spc_tev}
\end{eqnarray}
where $Q \equiv 4 \epsilon_{\rm IR} \Gamma$,
$q \equiv \Eg / Q(\Gamma-\Eg)$;
$\sgT$ is the Thomson cross section,
and $\Eg$ the energy of the upscattered photons in $m_{\rm e}c^2$ unit.
Substituting equation~(\ref{eq:IR_Crab}),
integrating $d^2 N/dtd\Eg$ over $\epsilon_{\rm IR}$,
and multiplying the $\gamma$-ray energy ($\Eg m_{\rm e}c^2$) and
the electron number ($N_{\rm e}$) in the gap,
we obtain the flux density of the upscattered, TeV photons 
as a function of $\Eg$.

We finally consider the extrinsic absorption of the TeV photons
outside of the gap.
For a homogeneous and isotropic IR field,
the optical depth becomes
\begin{equation}
  \tau(\Eg) 
  = \frac{\rlc}{2}
    \int_{\epsilon_{\rm IR,min}}^{\epsilon_{\rm IR,max}}
          \frac{dN_{\rm IR}}{d\epsilon_{\rm IR}}
          \sgP(\epsilon_{\rm IR},\Eg,\mu_{\rm c}) d\epsilon_{\rm IR},
  \label{eq:tauTeV}
\end{equation}
where the path length is assumed to be $\rlc/2$.

\begin{figure} 
\centerline{ \epsfxsize=8.5cm \epsfbox[200 20 500 250]{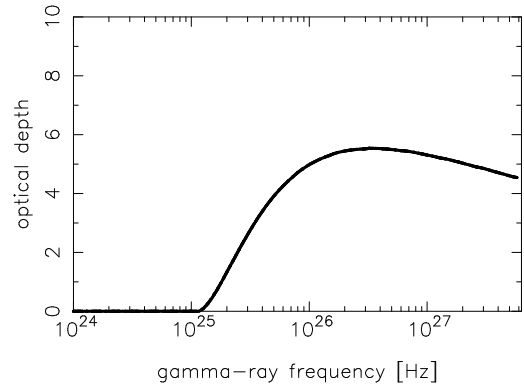} } 
\caption{\label{fig:Tau_Crab_45} 
Pair production optical depth for a TeV photon to be aborbed
in the homogeneous, isotropic IR field in the 
Crab-pulsar magnetosphere.
        }
\end{figure}

For the Crab pulsar,
$\tau \sim 5$ holds in TeV energies (fig.~\ref{fig:Tau_Crab_45}).
Therefore, the observed TeV flux
reduces to about $1\%$ of the intrinsic flux.

\section{Application to the Crab Pulsar}
\label{sec:results}

In this section, we apply the theory to the Crab pulsar.
The rotational frequency and the magnetic moment are
$\Omega= 188.1 \mbox{rad s}^{-1}$ and 
$\mu=3.38 \times 10^{30} \mbox{G cm}^3$.

\subsection{Electric Field Structure}
\label{sec:res_Ell}

HEAO 1 observations revealed that the
X-ray spectrum in the primary pulse phase 
is expressed by 
\begin{equation}
  \frac{dN_{\rm pl}}{d\epsilon_{\rm x}}
    = N_{\rm pl} \epsilon_{\rm x}{}^\alpha
  \quad (\epsilon_{\rm min} < \epsilon_{\rm x} < \epsilon_{\rm max}),
  \label{eq:dNdE_0}
\end{equation}
with $\alpha=-1.81$ 
and $N_{\rm pl}=5.3 \times 10^{15}d{}^2 (r_0/\rlc)^{-2}$
(Knight 1982),
where $d$ refers to the distance in kpc.
We adopt
$\epsilon_{\rm min}=0.1 \mbox{keV} / 511 \mbox{keV}$ and
$\epsilon_{\rm max}= 50 \mbox{keV} / 511 \mbox{keV}$.
Substituting this power-law spectrum into equation~(\ref{eq:def_etap_0}),
we can solve the Vlasov equations by the method described in \S~2.

We consider four representative boundary conditions:
We choose $(j_1,j_2)=(0,0)$, $(0.3,0)$, $(0.6,0)$, and $(0,0.3)$ 
as cases~1, 2, 3, and 4, respectively.
That is, for case~2 (or case~4), 
the positronic (or electronic) current density flowing into the gap
per unit flux tube at the inner (or outer) boundary is $30\%$ of 
the typical Goldreich-Julian value, $\Omega /2\pi$.
We fix $j_{\rm gap}= 0.01$ for all the four cases,
because the solution forms a \lq brim' 
to disappear (fig.~2 in Hirotani \& Okamoto 1998)
if $j_{\rm gap}$ exceeds a few percent.
In what follows, we adopt $45^\circ$ as the magnetic inclination.

\begin{figure} 
\centerline{ \epsfxsize=8.5cm \epsfbox[200 20 500 250]{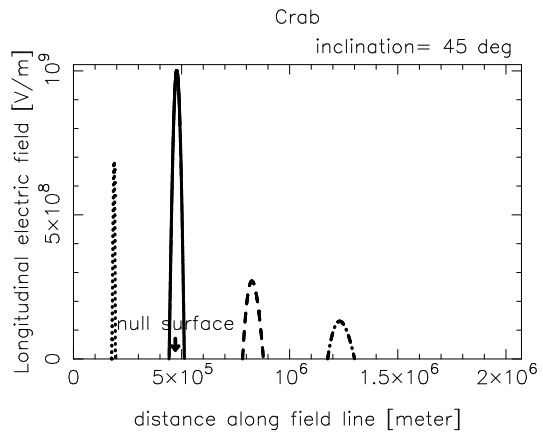} } 
\caption{\label{fig:Ell_Crab_45} 
Distribution of $\Ell(\xi)$.
The solid, dashed, dash-dotted, and dotted lines correspond to the
cases 1, 2, 3, and 4, respectively (see text).
        }
\end{figure}

The results of $\Ell(\xi)$ for the four cases are presented in
figure~\ref{fig:Ell_Crab_45}.
The abscissa designates the distance along the last-open field line
and covers the range from the neutron star surface ($s=0$)
to the position where the disance equals 
$s= 1.2 \rlc= 1.91 \times 10^6$~m.

The solid line (case~1) shows that the gap is located around the 
null surface. 
However, the gap shifts outwards as $j_1$ increases,
as the dashed (case~2) and dash-dotted (case~3) lines indicate.
This result is consistent with what was predicted in 
Shibata and Hirotani (2000) analytically.

On the other hand, when $j_2$ increases,
the gap shifts inwards
and the potential drop, $\Phi(s_2)$, reduces significantly.
For example, we obtain $\Phi(s_2)=7.1 \times 10^{12}$~V for case~4,
whereas $1.7 \times 10^{13}$~V for case~2. 
The reasons are sixfold:
\ $\bullet$~In a stationary gap, the pair production optical depth,
$W/\lambda_{\rm p}$, equals the ratio 
$N_\gamma (j_{\rm gap}$/$j_{\rm tot})$,
where $\lambda_{\rm p}$ and $N_\gamma$ refer 
to the pair production mean free path,
and the number of $\gamma$-rays emitted by a single particle,
respectively.
\ $\bullet$~The increased X-ray density at small radii
reduces $\lambda_{\rm p}$.
\ $\bullet$~The ratio $j_{\rm gap}$/$j_{\rm tot}$ decreases
as $j_2$ increases.
\ $\bullet$~As a result, $W$ decreases very rapidly with increasing $j_2$.
\ $\bullet$~Owing to the rapidly decreasing $W$, 
$\Phi(s_2)$ significantly decreases,
although the local $\rhoGJ$, and hence $d\Ell/ds$ increases 
at small radii.
\ $\bullet$~As $W$ decreases, $N_\gamma$ decreases to some extent;
however, this effect is passive and cannot change the conclusion.

\subsection{Gamma-ray Spectra}
\label{sec:res_spec}

\begin{figure} 
\centerline{ \epsfxsize=8.5cm \epsfbox[200 20 500 250]{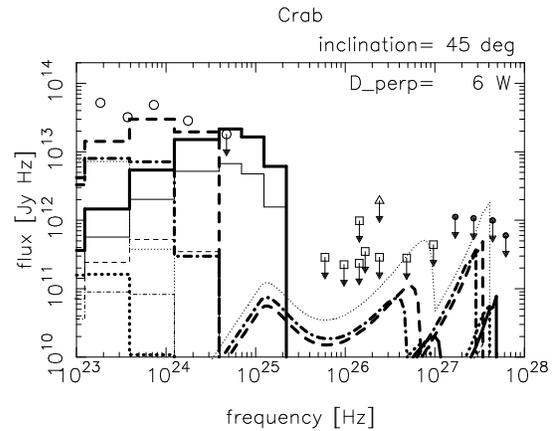} } 
\caption{\label{fig:Spc_Crab_45} 
Gamma-ray spectra from the Crab pulsar magnetosphere.
The thick (or thin) lines represent the flux of outwardly (or inwardly)
propagating $\gamma$-rays.
The solid, dashed, dash-dotted, and dotted lines correspond to 
the same cases as in figure~2.
        }
\end{figure}

The GeV spectra are readily computed from $g_+{}^i(\xi_2)$
and $g_-{}^i(\xi_1)$, while the TeV spectra are obtained 
by the method described in \S~\ref{sec:TeV_spc}.
We present the $\gamma$-ray spectra for the four cases
in figure~\ref{fig:Spc_Crab_45},
multiplying the cross sectional area of
$D_\perp{}^2= (6 W)^2$.
If $D_\perp$ increase twice, both the GeV and TeV fluxes 
increases four times.

In GeV energies,
the observational pulsed spectrum is obtained by
EGRET observations (filled circles; Nolan et al. 1993),
while in TeV energies,
only the upper limits are obtained 
by Whipple observations 
(open squares; Weekes et al. 1989; Reynolds et al. 1993;
 Goret et al. 1993; Lessard et al. 2000),
and by Durham observations (open triangle; Dowthwaite et al. 1984).
The filled circles denote the unpulsed flux obtained by 
CANGAROO observations (Tanimori et al. 1998).

The figure shows that the TeV fluxes are kept 
below the observational upper limits as a whole for appropriate
GeV fluxes.
Therefore, we can conclude that the problem of the excessive TeV flux
does not arise for a reasonable IR density for the Crab pulsar.

It is noteworthy that the GeV spectrum depends on $j_1$ and
$j_2$ significantly.
In particular, in case~4 (as the dotted lines show), 
the GeV emission significantly decreases and softens,
because both the potential drop and the maximum of $\Ell$ reduce
as the gap shifts inwards.
As a result, it becomes impossible to explain the EGRET flux
around $10^{24}$~Hz,
if the gap is located well inside of the null surface.

\section{Discussion}
\label{sec:discussion}

In summary, we have developed a one-dimensional model for 
an outer-gap accelerator in the magnetosphere of a rotation-powered pulsar.
When the electronic current flows into the gap
from the outer boundary,
the gap shifts inwards to emit very soft GeV emissions.
Applying this method to the Crab pulsar, we find that  
the gap should be located near to or outside of the null surface,
so that the observed spectrum of pulsed GeV fluxes may be emitted
via curvature process.
By virtue of the absorption by the dense IR field in the magnetosphere, 
the problem of excessive TeV emission does not arise.

Let us briefly compare the present method with ZC97,
who considered that the gap width is limited by the
surface X-rays due to the bombardment of the particles produced in the gap.
The magnetospheric X-rays considered in this paper
is much denser than the surface X-rays due to the bombardment. 
As a result, the localized gap in the present paper produces less
intrinsic TeV flux compared with what would be obtained in ZC97 picture.

For cases~1, 2, and 3, 
the intrinsic TeV luminosity is comparable or less than the GeV one.
Therefore, the Lorentz factors are limited primarily by curvature 
process (eq.[\ref{eq:terminal}]).
For case~4, however,
the intrinsic TeV luminosity well exceeds the GeV one;
therefore, the radiation-reaction forces are due to IC scatterings
rather than the curvature process.
In fact, we may expect a sufficient GeV flux via IC scatterings 
when the gap is located well inside of the null surface.
This is because the dense X-ray field will limit the particle Lorentz 
factors small (Paper~II),
and because the less-energetic particles scatter 
copious IR photons into lower $\gamma$-ray energies
with large cross sections ($\sim \sgT$).
There is room for further investigation on this issue.

\par
\vspace{1pc}\par

One of the authors (K. H.) wishes to express his gratitude to
Drs. Y. Saito and A. K. Harding for valuable advice. 
He also thanks the Astronomical Data Analysis Center of
National Astronomical Observatory, Japan for the use of workstations.

\end{document}